\documentclass[final]{aipproc}
\layoutstyle{8x11single}

\begin{document}

\author{A. M\'esz\'aros}{
  address={Charles University, Prague, Czech Republic},
  email={meszaros@cesnet.cz},
}

\author{ L.G. Bal\'azs}{
  address={Konkoly Observatory, Budapest, Hungary},
  email={balazs@konkoly.hu},
}

\author{Z. Bagoly}{
  address={E\"otv\"os University, Budapest, Hungary},
  email={zsolt@yela.elte.hu},
}

\author{P. Veres}{
  address={E\"otv\"os University, Budapest, Hungary},
  email={veresp@gmail.com},
}

\title{
 Anisotropy in the sky distributions of the short and intermediate
gamma-ray bursts: Breakdown of the cosmological principle?}
\date{2008/12/08}


\begin{abstract}
After the discovery of the anisotropy in the
sky-distribution of intermediate gamma-ray bursts recently also 
the distribution of the short 
gamma-ray bursts is proven to be anisotropic. The impact of these behaviors 
on the validity of the cosmological principle is shortly discussed.
\end{abstract}

\maketitle

\section{INTRODUCTION}

The cosmological principle requires that the Universe be
spatially homogeneous and isotropic in average on scales larger than the 
size of any structure  \cite{Peebles}. Observations show that the 
greatest structures (filaments, voids, superclusters, ...) have sizes
around 400 Mpc \cite{Peacock}.
Hence, at redshifts $z < 0.1$ the matter distribution in
the Universe is anisotropic and inhomogeneous. There are
further supports (both observational \cite{bro,pa92,holba,Rudnick} 
and theoretical \cite{hjk99,spri} ones)
that such structures  exist also at redshift $z < 1$.
If the structures with sizes $\simeq 400$ Mpc are not
distributed isotropically at $z <1$, then some structures should exist
even with sizes $\simeq (0.4-3.3)$ Gpc. In other words, if this is
the case, then the scale - where the
averaging should be done - should be at least of order $\sim 1\, Gpc$. 
In this case the fulfilement of the cosmological principle would be in 
doubt.

The angular
distribution of the gamma-ray bursts (GRBs) allows to test this 
principle
too, because - if this principle holds - GRBs should be distributed
isotropically on the sky, if they are dominantly at $ z > 0.1$. For this
test it is convenient that GRBs are well
seen in the gamma-band also in the Galactic plane.

At the last years the authors provided several
different tests probing the intrinsic isotropy in the angular
sky-distribution of GRBs collected in BATSE Catalog (for the survey and
for further details see \cite{Vavrek}).
Shortly summarizing the results of these studies one may conclude:
A. The long subgroup ($T_{90} > 10\; s$) seems to be distributed
isotropically; B. The intermediate subgroup ($2\;s \le T_{90}
\le 10\;s $) is distributed anisotropically on the $\simeq
(96-97)$\% significance level; C. For the short subgroup ($2\; s >
T_{90}$) the assumption of isotropy is rejected only on the $92$\%
significance level; D. The long and the short subclasses,
respectively, are distributed differently on the $99.3$\%
significance level. (About the definition of subclasses and other 
relevant topics see, e.g., \cite{ho98,mu98,ho00,li01,ho02,ho08};
$T_{90}$ is the duration of a GRB, during which time
the $90$\% of the radiated energy is received.)

\section{DESCRIPTION OF THE NEWEST TESTS}

Because mainly the situation concerning the short GRBs was unclear, we
provided a new more powerful testing of the isotropy on the BATSE data. 
We used three methods.  (The detailed description of these three methods 
with references is given in  \cite{Vavrek}.)

The first one is the method called {\bf Voronoi tesselation (VT)}. The
Voronoi diagram - also known as Dirichlet
tesselation or Thiessen polygons - is a fundamental structure in
computational geometry. Generally, this diagram provides a
partition of a point pattern ("point field", also "point process")
according to its spatial structure, which can be used for
analyzing  the underlying point process. The points on sphere may be
distributed  completely randomly or
non-randomly; the non-random distribution may have different
characters (clustering, filaments, etc.).

The second method is called {\bf Minimal spanning tree (MST)}.
Contrary to VT, this method considers the distances (edges) among the
points (vertices). Clearly, there are $N(N-1)/2$ distances among
$N$ points. A spanning tree is a system of lines connecting all
the points without any loops. The MST is a
system of connecting lines, where the sum of the lengths is
minimal among all the possible connections between the points.
In our study the spherical version of MSF was used.
The $N-1$ separate connecting lines (edges) together define the
minimal spanning tree. The statistics of the lengths and the
$\alpha_{MST}$ angles between the edges at the vertices  can be
used for testing the randomness of the point pattern. 

The third method uses the {\bf Multifractal spectrum (MS)}.
The idea here is the following: Let denote $P(\varepsilon)$ the
probability for
finding a point in an area of $\varepsilon$ radius.  If
$P(\varepsilon)$ scales as $\varepsilon^{\alpha}$ (i.e.
$P(\varepsilon)\propto \varepsilon^{\alpha}$), then $\alpha$ is
called the local fractal dimension (e.g. $\alpha=2$ for a
completely random process on the plane). In the case of a
monofractal $\alpha$ is independent on the position. A
multifractal (MFR) on a point process can be defined as
unification of the subsets of different (fractal) dimensions.
One usually denotes with $f(\alpha)$ the fractal
dimension of the subset of points at which the local fractal
dimension is in the interval  of $\alpha,\alpha+d\alpha$. The
contribution of these subsets to the whole pattern is not
necessarily equally weighted, practically it depends on the
relative abundances of subsets. The $f(\alpha)$ functional
relationship between the fractal dimension of subsets and the
corresponding local fractal dimension is called the MFR or
Hausdorff spectrum.

 The randomness of the point field on the sphere
 can be tested with respect to different criteria. Since
different non-random behaviors are sensitive for different types
of criteria of non-randomness, it is not necessary that  all
possible tests using different measures reject the assumption of
randomness. We defined several test-variables
which are sensitive to different stochastic properties of the
underlying point pattern. For VT
any of the four quantities characterizing the Voronoi cell - i.e.
the area, the perimeter, the number of vertices, and the inner
angles - can be used as test-variables or even some of their
combinations, too. We used three quantities obtained for MST:
variance of the MST edge-length
($\sigma(L_{MST})$); mean MST edge-length ($L_{MST}$);
and the mean angle between edges ($\alpha_{MST}$).
For the multifractal spectrum
the only used variable is the $f(\alpha)$ multifractal spectrum,
which is a sensitive tool for testing the non-randomness of a point
pattern. Together we had 13 variables, not being independent. The
significance levels were calculated by Monte Carlo simulations.

 \begin{center}
\includegraphics[width= {0.52\columnwidth} ]{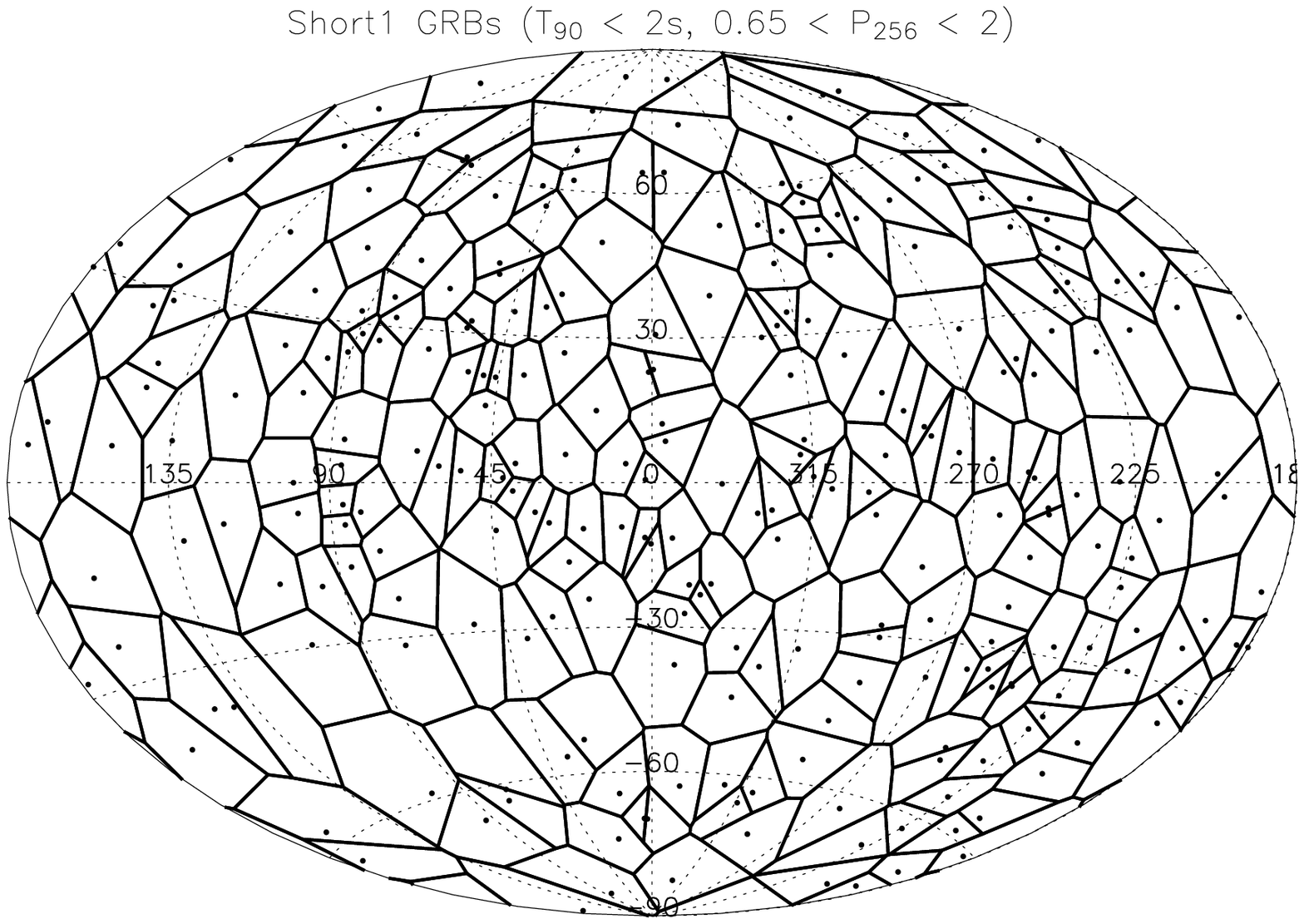}\\
\end{center}

\section{THE SAMPLES}

We  divided the BATSE's GRBs into
three groups: $short$ ($T_{90} < 2s$), $intermediate$ ($2s\leq
T_{90} \leq 10s$) and $long$ ($T_{90} > 10s$). To avoid the
problems with the changing detection threshold we omitted the GRBs
having a  peak flux $P_{256} \leq 0.65 \ photons \ cm^{-2} \
s^{-1}$. The bursts
may emerge at very different distances in the line of sight and it
may happen that the stochastic structure of the angular
distribution depends on it. Therefore, we also made tests on the
bursts with $P_{256} < 2\ photons \ cm^{-2} \
s^{-1}$ in the short and long population, separately.
All this means that we studied five samples:
"Short1" with $T_{90}<2$ s and  $0.65 < P_{256} < 2$ containing 261 
GRBs;
"Short2" with $T_{90}<2$ s and $0.65 < P_{256}$ containing 406 GRBs;
"Intermediate" with $2\;s \leq T_{90} \leq 10\;s $ and $0.65 < P_{256}$
containing 253 GRBs;
 "Long1"  with  $T_{90}>2$ s and $0.65 < P_{256} < 2$ containing 676 
GRBs;
and "Long2" with  $T_{90}>10$ s and $0.65 < P_{256}$  containing 966 
GRBs.
VT of the Short1 sample in Galactical coordinates is shown on the 
Figure.

\section{THE RESULTS}

{\bf Both the Short1 and the Short2 samples deviate
significantly on the 99.90\% and  99.98\% significance levels
from the full randomness. Also the Intermediate sample gave
a significant deviation (98.51\%) from the full randomness in accordance
with the earlier study \cite{Mesz}. The long samples remained random.}

\section{DISCUSSION AND CONCLUSION}

The subgroups of GRBs, detected by BATSE, show different sky
distribution: Both the short and intermediate ones are distributed
anisotropically on a high significance level; the long ones seem to be
distributed isotropically. The situation concerning the intermediate 
ones
is unclear (Is it a physically different subgroup or not?), but there is
no doubt that the short and long ones are
different phenomena.  Because the short ones are distributed
anisotropically (in addition - on large angular scales \cite{mesz00}), 
{\bf up to the redshifts - where the short ones dominate - the 
cosmological principle hardly can be fulfilled.} The directly measured
redshifts of the short GRBs - detected by Swift \cite{swift} - suggest 
that just this happens up to $z < 1$ (see Table 1 of \cite{Kann}).
It is not sure that the short GRBs detected by
BATSE, and the short GRBs detected by Swift together with measured
redshifts are at the same redshift range. But, even keeping in mind this
eventuality, it is well possible {\bf that the validity of the
cosmological principle is in doubt at $z <1$. A support for this point 
of view was given by \cite{Vavrek}.}

Note here that a similar sceptical points of view about the validity
of the cosmological principle were claimed, too, from the studies of  
fully different topics \cite{Collins,birch,kash}.

\section{ACKNOWLEDGEMENTS}

 This research was supported from
OTKA grants T048870 and T077795, by a GAUK grant No.46307,
by the Research Program MSM0021620860 of the
Ministry of Education of the Czech Republic (A.M.), and by a Pol\'anyi
grant KFKT-2006-01-00012 (P.V.).

\end{document}